\documentclass[ALICE,manyauthors]{cernphprep}
\usepackage[comma,square,numbers,sort&compress]{natbib}
\usepackage{hyperref}
\usepackage{lineno}
\usepackage{xspace}
\usepackage{xcolor} 
\usepackage{csquotes}
\usepackage{multirow}
\usepackage{upgreek}
\usepackage{booktabs}
\usepackage[T1]{fontenc}
\usepackage{orcidlink}

\begin{document}
%

\newcommand{\pp}           {pp\xspace}
\newcommand{\ppbar}        {\ensuremath{\mathrm{p\overline{p}}}\xspace} 
\newcommand{\ee}           {\ensuremath{\mathrm{e^{+}e^{-}}}\xspace} 
\newcommand{\ep}           {\ensuremath{\mathrm{e^\pm p}}\xspace} 
\newcommand{\XeXe}         {\mbox{Xe--Xe}\xspace}
\newcommand{\PbPb}         {\mbox{Pb--Pb}\xspace}
\newcommand{\pA}           {\mbox{pA}\xspace}
\newcommand{\pPb}          {\mbox{p--Pb}\xspace}
\newcommand{\AuAu}         {\mbox{Au--Au}\xspace}
\newcommand{\dAu}          {\mbox{d--Au}\xspace}

\newcommand{\s}            {\ensuremath{\sqrt{s}}\xspace}
\newcommand{\snn}          {\ensuremath{\sqrt{s_{\mathrm{NN}}}}\xspace}
\newcommand{\pt}           {\ensuremath{p_\mathrm{T}}\xspace}
\newcommand{\kt}           {\ensuremath{k_\mathrm{T}}\xspace}
\newcommand{\avdndeta}     {\ensuremath{\langle\mathrm{d}N_\mathrm{ch}/\mathrm{d}\eta\rangle_{|\eta|<0.5}}\xspace}
\newcommand{\fprompt}      {\ensuremath{f_\mathrm{prompt}}\xspace}
\newcommand{\eff}          {\ensuremath{(\mathrm{Acc}\times\epsilon)}\xspace}
\newcommand{\effprompt}    {\ensuremath{(\mathrm{Acc}\times\epsilon)_\mathrm{prompt}}\xspace}
\newcommand{\INEL}         {\ensuremath{\mathrm{INEL} > 0}\xspace}

\newcommand{\LT}           {L{\'e}vy-Tsallis\xspace}
\newcommand{\TeV}          {\ensuremath{\mathrm{TeV}}\xspace}
\newcommand{\GeV}          {\ensuremath{\mathrm{GeV}}\xspace}
\newcommand{\MeV}          {\ensuremath{\mathrm{MeV}}\xspace}
\newcommand{\GeVc}         {\ensuremath{\mathrm{GeV}/c}\xspace}
\newcommand{\MeVc}         {\ensuremath{\mathrm{MeV}/c}\xspace}
\newcommand{\GeVmass}      {\ensuremath{\mathrm{GeV}/c^2}\xspace}
\newcommand{\MeVmass}      {\ensuremath{\mathrm{MeV}/c^2}\xspace}
\newcommand{\lumi}         {\ensuremath{\mathcal{L}}\xspace}
\newcommand{\fm}           {\ensuremath{\mathrm{fm}}\xspace}
\newcommand{\fmc}          {\ensuremath{\fm/c}\xspace}
\newcommand{\ctau}         {\ensuremath{c\tau}\xspace}

\newcommand{\ITS}          {\ensuremath{\mathrm{ITS}}\xspace}
\newcommand{\TOF}          {\ensuremath{\mathrm{TOF}}\xspace}
\newcommand{\ZDC}          {\ensuremath{\mathrm{ZDC}}\xspace}
\newcommand{\ZDCs}         {\ensuremath{\mathrm{ZDCs}}\xspace}
\newcommand{\ZNA}          {\ensuremath{\mathrm{ZNA}}\xspace}
\newcommand{\ZNC}          {\ensuremath{\mathrm{ZNC}}\xspace}
\newcommand{\SPD}          {\ensuremath{\mathrm{SPD}}\xspace}
\newcommand{\SDD}          {\ensuremath{\mathrm{SDD}}\xspace}
\newcommand{\SSD}          {\ensuremath{\mathrm{SSD}}\xspace}
\newcommand{\TPC}          {\ensuremath{\mathrm{TPC}}\xspace}
\newcommand{\TRD}          {\ensuremath{\mathrm{TRD}}\xspace}
\newcommand{\VZERO}        {\ensuremath{\mathrm{V0}}\xspace}
\newcommand{\VZEROA}       {\ensuremath{\mathrm{V0A}}\xspace}
\newcommand{\VZEROC}       {\ensuremath{\mathrm{V0C}}\xspace}
\newcommand{\Vdecay} 	     {\ensuremath{V^{0}}\xspace}

\newcommand{\pip}          {\ensuremath{\uppi^{+}}\xspace}
\newcommand{\pim}          {\ensuremath{\uppi^{-}}\xspace}
\newcommand{\kap}          {\ensuremath{\mathrm{K}^{+}}\xspace}
\newcommand{\kam}          {\ensuremath{\mathrm{K}^{-}}\xspace}
\newcommand{\pbar}         {\ensuremath{\mathrm\overline{p}}\xspace}
\newcommand{\kzero}        {\ensuremath{\mathrm{K}^{0}_\mathrm{S}}\xspace}
\newcommand{\lmb}          {\ensuremath{\Lambda}\xspace}
\newcommand{\almb}         {\ensuremath{\overline{\Lambda}}\xspace}
\newcommand{\Om}           {\ensuremath{\Omega^-}\xspace}
\newcommand{\Mo}           {\ensuremath{\overline{\Omega}^+}\xspace}
\newcommand{\X}            {\ensuremath{\Xi^-}\xspace}
\newcommand{\Ix}           {\ensuremath{\overline{\Xi}^+}\xspace}
\newcommand{\Xis}          {\ensuremath{\Xi^{\pm}}\xspace}
\newcommand{\Oms}          {\ensuremath{\Omega^{\pm}}\xspace}
\newcommand{\degree}       {\ensuremath{^\mathrm{o}}\xspace}

\newcommand{\OmegacZero}     {\ensuremath{\Omega_\mathrm{c}^{0}}\xspace}
\newcommand{\XicZero}        {\ensuremath{\Xi_\mathrm{c}^{0}}\xspace}
\newcommand{\XicPlus}        {\ensuremath{\Xi_\mathrm{c}^{+}}\xspace}
\newcommand{\XicZeroPlus}    {\ensuremath{\Xi_\mathrm{c}^{0,+}}\xspace}
\newcommand{\Sigmac}         {\ensuremath{\Sigma_\mathrm{c}^{0,++}}\xspace}
\newcommand{\Lambdac}        {\ensuremath{\Lambda_\mathrm{c}^{+}}\xspace}
\newcommand{\JPsi}           {\ensuremath{\mathrm{J}/\uppsi}\xspace}
\newcommand{\acceff}         {\enusremath{\mathrm{Acc.}\times\varepsilon}\xspace}
\newcommand{\Dzero}          {\ensuremath{\mathrm{D}^{0}}\xspace}
\newcommand{\Dplus}          {\ensuremath{\mathrm{D}^{+}}\xspace}
\newcommand{\Dstar}          {\ensuremath{\mathrm{D}^{*+}}\xspace}
\newcommand{\Ds}             {\ensuremath{\mathrm{D_{s}^{+}}}\xspace}
\newcommand{\DstarNoSign}    {\ensuremath{\mathrm{D}^{*}}\xspace}
\newcommand{\DzerotoKpi}     {\ensuremath{\Dzero\to\mathrm{K}^-\uppi^+}\xspace}
\newcommand{\DplusToKpipi}   {\ensuremath{\Dplus\to\mathrm{K}^-\uppi^+\uppi^+}\xspace}
\newcommand{\DstarToDzeropi} {\ensuremath{\Dstar\to\Dzero(\to\mathrm{K}^-\uppi^+)\uppi^+}\xspace}
\newcommand{\DsToPhipiToKKpi}{\ensuremath{\Ds\to\upphi\uppi^+\to\mathrm{K}^+\mathrm{K}^-\uppi^+}\xspace}
\newcommand{\ccbar}          {\ensuremath{\mathrm{c\overline{c}}}\xspace}
\newcommand{\bbbar}          {\ensuremath{\mathrm{b\overline{b}}}\xspace}
\newcommand{\DsonetoDstarKs} {\ensuremath{\Dsone\to\Dstar\kzero}\xspace}
\newcommand{\DstwotoDplusKs} {\ensuremath{\Dstwo\to\Dplus\kzero}\xspace}
\newcommand{\Dsone}          {\ensuremath{\mathrm{D_{s1}^+}}\xspace}
\newcommand{\Dstwo}          {\ensuremath{\mathrm{D_{s2}^{*+}}}\xspace}
\newcommand{\DsonePrecise}   {\ensuremath{\mathrm{D_{s1}}(1^{+})(2536)^+}\xspace}
\newcommand{\DstwoPrecise}   {\ensuremath{\mathrm{D_{s2}^{*}(2^{+})(2573)^+}}\xspace}
\newcommand{\PhiPrecise}     {\ensuremath{\mathrm{\upphi(1020)}}\xspace}
\newcommand{\Hc}             {\ensuremath{\mathrm{H_c}}\xspace}
\newcommand{\fcToHc}         {\ensuremath{f(\mathrm{c}\to\Hc)}\xspace}
\newcommand{\KstarPlus}      {\ensuremath{\mathrm{K}^{*}(892)^+}\xspace}
\newcommand{\KstarZero}      {\ensuremath{\mathrm{K}^{*}(892)^0}\xspace}
\newcommand{\Kstar}          {\ensuremath{\mathrm{K}^*}\xspace}
\newcommand{\KzeroTopipi}    {\ensuremath{\kzero\to\uppi^+\uppi^-}\xspace}

\renewcommand{\pi}{\uppi}
\renewcommand{\phi}{\upphi}

\newcommand{\pythia}[1] {\textsc{PYTHIA}\;#1}
\newcommand{\shm}       {\textsc{SHM}}
\newcommand{\shmc}      {\textsc{SHMc}}
\newcommand{\fonll}     {\textsc{fonll}}
\newcommand{\gmvfns}    {\textsc{gm-vfns}}

\begin{titlepage}
\PHyear{2024}       
\PHnumber{233}      
\PHdate{10 September}  

\title{First measurement of $\pmb{\DsonePrecise}$ and $\pmb{\DstwoPrecise}$ production in proton--proton collisions at $\pmb{\s=13~\TeV}$ at the LHC}
\ShortTitle{$\Dsone$ and $\Dstwo$ production}   

\Collaboration{ALICE Collaboration\thanks{See Appendix~\ref{app:collab} for the list of collaboration members}}
\ShortAuthor{ALICE Collaboration} 

\begin{abstract}
The production yields of the orbitally excited charm-strange mesons \DsonePrecise and \DstwoPrecise were measured for the first time in proton--proton (pp) collisions at a center-of-mass energy of \mbox{$\s=13~\TeV$} with the ALICE experiment at the LHC. The \Dsone and \Dstwo mesons were measured at midrapidity ($|y|<0.5$)
in minimum-bias and high-multiplicity pp collisions in the transverse-momentum interval $2<\pt<24~\GeVc$. Their production yields relative to the \Ds ground-state yield were found to be compatible between minimum-bias and high-multiplicity collisions, as well as with previous measurements in \ep and \ee collisions. The measured $\Dsone/\Ds$ and $\Dstwo/\Ds$ yield ratios are described by statistical hadronization models and can be used to tune the parameters governing the production of excited charm-strange hadrons in Monte Carlo generators, such as \pythia{8}.
\end{abstract}
\end{titlepage}

\setcounter{page}{2} 


The production of charm mesons with spin zero (D mesons) or one (\DstarNoSign mesons) and orbital angular momentum $L=0$  has been extensively studied in recent years in proton--proton (pp) collisions at the LHC by the ALICE~\cite{ALICE:2019nxm, ALICE:2021mgk, ALICE:2023sgl}, ATLAS~\cite{ATLAS:2015igt}, CMS~\cite{CMS:2021lab}, and LHCb~\cite{LHCb:2015swx, LHCb:2016ikn} Collaborations. The production cross sections are generally described within uncertainties by perturbative QCD calculations at next-to-leading order (NLO) with next-to-leading log resummation (e.g.,~FONLL~\cite{Cacciari:1998it, Cacciari:2001td, Cacciari:2012ny} and GM-VFNS~\cite{Kniehl:2012ti, Helenius:2018uul}) via the factorization of three terms, namely the parton distribution functions (PDFs) of the incoming protons, the partonic cross section, and the fragmentation functions (FFs) describing the transition from charm quarks to the final hadrons. In these calculations, the FFs are typically parameterized from measurements in \ee and \ep collisions~\cite{Braaten:1994bz} under the assumption that the hadronization of charm quarks into charm hadrons is a universal process, independent of the collision system. The relative abundances of the different D-meson species were found to be compatible with those measured in \ee and \ep collisions~\cite{ALICE:2021mgk, ALICE:2023sgl}. A significant discrepancy was instead observed at midrapidity for the charm baryons, whose production relative to the one of the \Dzero meson turned out to be enhanced in pp collisions compared to \ee and \ep collisions~\cite{ALICE:2020wfu,ALICE:2020wla,Acharya:2021vpo,Acharya:2021dsq,Acharya:2021vjp,ALICE:2022cop,CMS:2019uws,CMS:2023frs}, implying a modification of the fragmentation fractions of charm quarks to the various charm-hadron species in pp collisions at LHC energies compared to those measured at lepton colliders at the LHC~\cite{ALICE:2021dhb,ALICE:2023sgl}. 

The charm-meson spectroscopy has also progressed significantly in the last few decades, with also the discovery of several excited charm and charm-strange states and the determination of their properties~\cite{CLEO:2003ggt,BaBar:2006gme,BaBar:2010zpy,Belle:2007hht,BaBar:2014jjr,LHCb:2012uts,LHCb:2013jjb,LHCb:2016lxy,LHCb:2015klp}. However, the production yields of charm resonances were only measured at \ee and \ep colliders~\cite{OPAL:1997ekz,ALEPH:2001fud,ZEUS:2008nzg,ZEUS:2012gyr}, and no experimental result is available in hadronic collisions. The knowledge about the production yields of such states would provide important information about the hadronization of charm quarks produced in hadronic collisions, since they contribute to the ground-state charm-hadron yields via strong decays. This is, for example, the case in models based on statistical hadronization, in which the yields of the various charm-hadron species are assumed to follow the relative thermal densities and hence depend on the state mass and spin-degeneracy factor $2J+1$, where $J$ is the total angular momentum~\cite{Andronic:2003zv,He:2019tik}. Charm resonances are typically not included in Monte Carlo (MC) generators, such as PYTHIA~8~\cite{Skands:2014pea}, due to the lack of knowledge about their production and decays. Moreover, the production of short-lived resonances is important to study the hadronic phase of the system created in heavy-ion collisions.
In the case when a resonance has a lifetime comparable to that of the hadronic phase, suppression or regeneration of the resonance state due to interactions of its decay products with the hadron gas is observed~\cite{ALICE:2019xyr,ALICE:2019etb,ALICE:2021xyh,ALICE:2023ifn}. A lower limit of the lifetime of the hadronic phase in central Pb--Pb collisions of about 4--7 \fmc was determined via the measurement of the \pt-integrated yield ratio of the \KstarZero resonance ($\ctau\approx 4~\mathrm{fm}$) to K$^\pm$ mesons~\cite{ALICE:2019xyr}. In addition, a hint of suppression was also measured in \pPb and high-multiplicity pp collisions, suggesting the possible presence of rescattering effects and, thus, of a hadronic phase with a short but non-zero lifetime in small collision systems~\cite{ALICE:2019etb}.

In this letter, the first measurement of the production yields of the orbitally excited charm-strange mesons \DsonePrecise ($\ctau \approx 214~\mathrm{fm}$) and \DstwoPrecise ($\ctau \approx 11.7~\mathrm{fm}$)~\cite{ParticleDataGroup:2022pth} and their charge conjugates in minimum-bias and high-multiplicity pp collisions at a center-of-mass energy of $\s=13~\TeV$ is reported. In the following, \Dsone denotes \DsonePrecise, and \Dstwo stands for \DstwoPrecise.
The results, which are integrated in the transverse-momentum interval $2<\pt<24~\GeVc$ where it was possible to observe a stable signal, are divided by the production yield of the ground state \Ds and compared with predictions obtained with the statistical hadronization model. The measured ratios are also used to constrain the parameters in PYTHIA~8, which regulate the production of pseudovector and tensor charm mesons, both with the inclusion of rescattering effects in the hadronic phase or without them. Finally, the excited-to-ground state yield ratios are exploited to compute the fragmentation fractions of charm quarks into the \Dsone and \Dstwo states. The measured values are compared to those obtained in \ee collisions.

The apparatus of the ALICE experiment and its performance during the Run 2 data-taking period are described in detail in Refs.~\citenum{ALICE:2008ngc} and \citenum{ALICE:2014sbx}.
The main sub-detectors, located at midrapidity ($|y| < 0.5$), employed to perform the measurements presented in this letter are: the Inner Tracking System (ITS), for tracking and vertex reconstruction; the Time Projection Chamber (TPC), for tracking and particle identification (PID); the Time-Of-Flight (TOF) detector, for particle identification. The \VZERO detector, composed of two arrays of scintillators located on both sides of the collision region ($-3.7<\eta<-1.7$ and $2.8<\eta<5.1$),  is used for trigger purposes as well as to measure the event multiplicity~\cite{ALICE:2020swj}. The latter is determined from the percentile distribution of the summed amplitude recorded by both sides of the \VZERO detector, the V0M. Percentile values for higher multiplicity collisions are close to 0\% and for lower ones close to 100\%. In the following, the $0$--$0.1\%$ V0M multiplicity class is denoted as HMV0.

The measurements reported in this letter were performed on the sample of \pp collisions \mbox{at $\s=13$}~TeV collected with the ALICE experiment from 2016 to 2018. The data were recorded using a minimum-bias trigger (MB) requiring coincident signals in both \VZERO scintillator arrays. Offline selection criteria were applied to remove beam-induced background events, exploiting the timing information from the \VZERO arrays and the correlation between the number of clusters and track segments reconstructed in the two innermost layers of the ITS. Events with pileup of collisions within the same bunch crossing, with an estimated probability ranging from $10^{-3}$ to $10^{-2}$  depending on the beam conditions, were excluded by rejecting events with more than one reconstructed primary vertex~\cite{ALICE:2020swj}. To ensure uniform pseudorapidity acceptance, only events with a primary vertex position within $\pm 10$~cm from the nominal center of the apparatus along the beam direction were considered. Furthermore, the events satisfying the aforementioned selection criteria were required to possess at least one reconstructed track segment between the first two layers of the ITS within $|\eta| < 1$ (\INEL event class).

The resulting data sample consisted of about $1.8 \times 10^{9}$ \INEL and $0.3 \times 10^{9}$ HMV0 events, corresponding to integrated luminosities of about 32~$\text{nb}^{-1}$ and 7.7~$\text{pb}^{-1}$~\cite{ALICE-PUBLIC-2021-005}, respectively. The multiplicity percentile measured by the \VZERO detector was converted into an average charged-particle multiplicity, \avdndeta, by following the prescription detailed in Ref.~\citenum{ALICE:2020swj}. A trigger correction was applied to account for those events that fulfill the \INEL requirement but were not selected by the trigger. This correction factor $\epsilon^{\rm INEL} = 0.920 \pm 0.003$ was estimated with a detailed Monte Carlo simulation based on the PYTHIA~8 generator~\cite{Skands:2014pea} and the GEANT3 transport package~\cite{Brun:1082634}. For the HMV0 events the trigger was fully efficient and, therefore, a correction is not necessary~\cite{ALICE:2021npz}.

The \Dsone and \Dstwo mesons and their charge conjugates were measured at midrapidity through the hadronic decay channels \DsonetoDstarKs and \DstwotoDplusKs, whose branching ratios (BRs) are not yet measured~\cite{ParticleDataGroup:2022pth}. The \kzero mesons were reconstructed via \KzeroTopipi decays with a BR of $(69.20 \pm 0.05)\%$~\cite{ParticleDataGroup:2022pth}. \Dplus and \Dstar mesons were reconstructed in the decay channels \DplusToKpipi with a BR of $(9.38 \pm 0.16)\%$~\cite{ParticleDataGroup:2022pth} and \DstarToDzeropi with a BR of $(2.67 \pm 0.03)\%$~\cite{ParticleDataGroup:2022pth}, respectively.
The reconstruction and selection of \kzero-meson candidates closely followed the approaches presented in previous publications~\cite{ALICE:2020jsh,ALICE:2010vtz}. Pairs of tracks with opposite charge signs, with $|\eta| < 0.8$ and satisfying the track-quality and particle-identification (PID) criteria reported in Ref.~\citenum{ALICE:2020jsh}, were formed. Further selections based on the characteristic weak-decay topology of \kzero mesons were applied to reduce the combinatorial-background contribution.
Similarly, \Dzero- and \Dplus-meson candidates were obtained from pairs and triplets of tracks, respectively, with the proper charge signs and $|\eta| < 0.8$. The \Dstar-meson candidates were reconstructed by combining \Dzero candidates with tracks identified as pions and having $\pt > 50~\MeVc$. The signal selection exploited the reconstruction of decay-vertex topologies of D mesons displaced from the interaction vertex. A machine-learning approach based on Boosted Decision Trees (BDTs)~\cite{,10.1145/2939672.2939785,barioglio_luca_2022_7014886} was used to enhance the rejection of the combinatorial background and to separate D mesons produced directly in the charm-quark hadronization or through decays of excited charm-hadron states (prompt) from those originating from beauty-hadron decays (non-prompt). The quantities provided as input to the BDTs were based on the topological and kinematic properties of the D-meson candidates, and the PID information of their daughter tracks. The selection procedure and criteria were the same as those used in Refs.~\citenum{ALICE:2023brx} and~\citenum{ALICE:2022byg}, for \Dplus and \Dstar mesons respectively.
\Dsone- and \Dstwo-meson candidates were reconstructed by combining \Dstar and \Dplus mesons with \kzero mesons satisfying the aforementioned selection criteria. Furthermore, only \Dstar, \Dplus, and \kzero mesons with a reconstructed invariant mass within a window of $\pm 3\sigma$ around the reconstructed mass value were considered, where the signal-peak width ($\sigma$) and mean for the three meson species were estimated by fitting their particle-candidate invariant-mass distributions with a Gaussian plus a function to describe the combinatorial background. No additional requirements were applied to the orbitally excited charm-strange meson candidates.

\begin{figure}[tb]
  \centering
  \includegraphics[width = 0.4\textwidth]{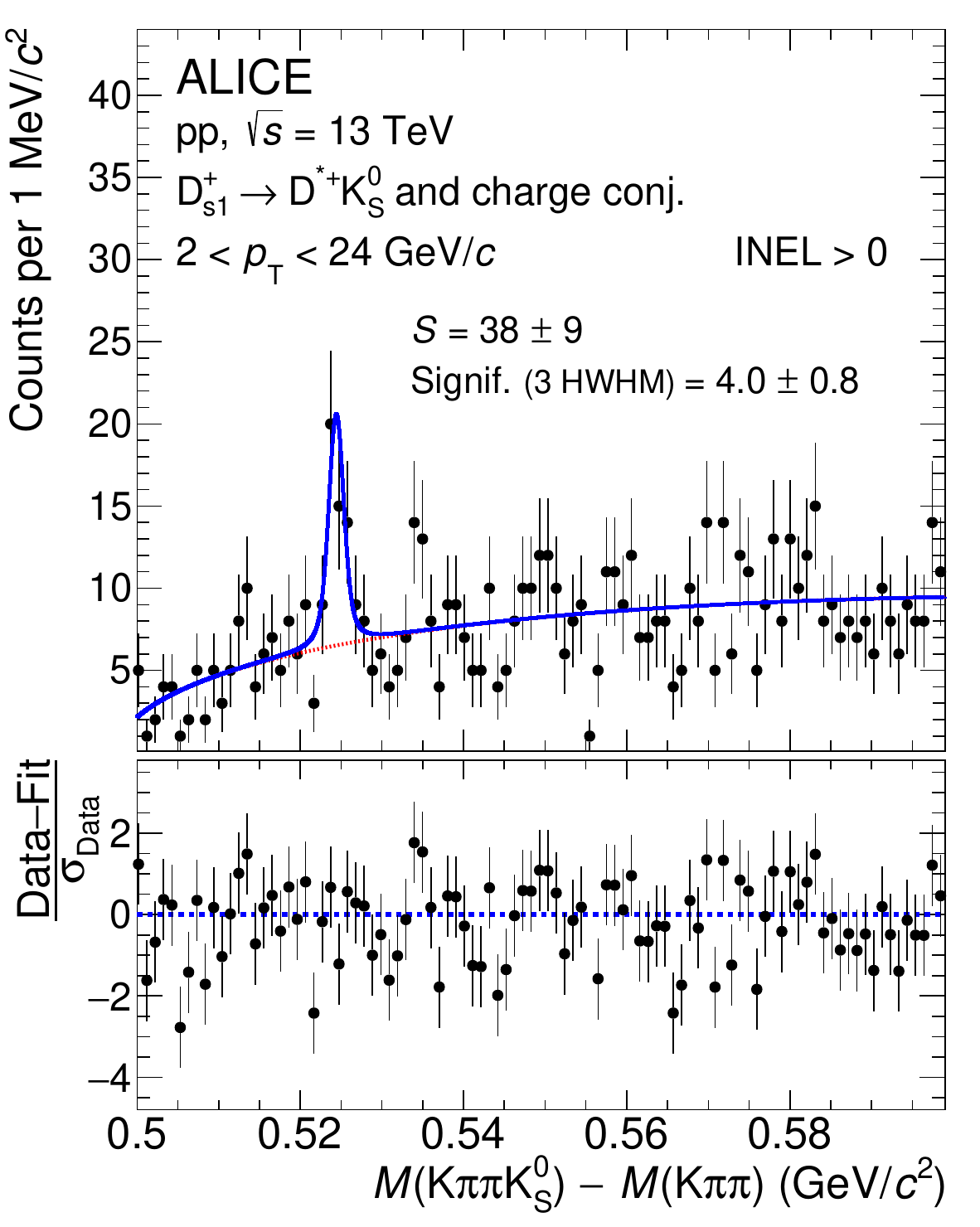}
  \includegraphics[width = 0.4\textwidth]{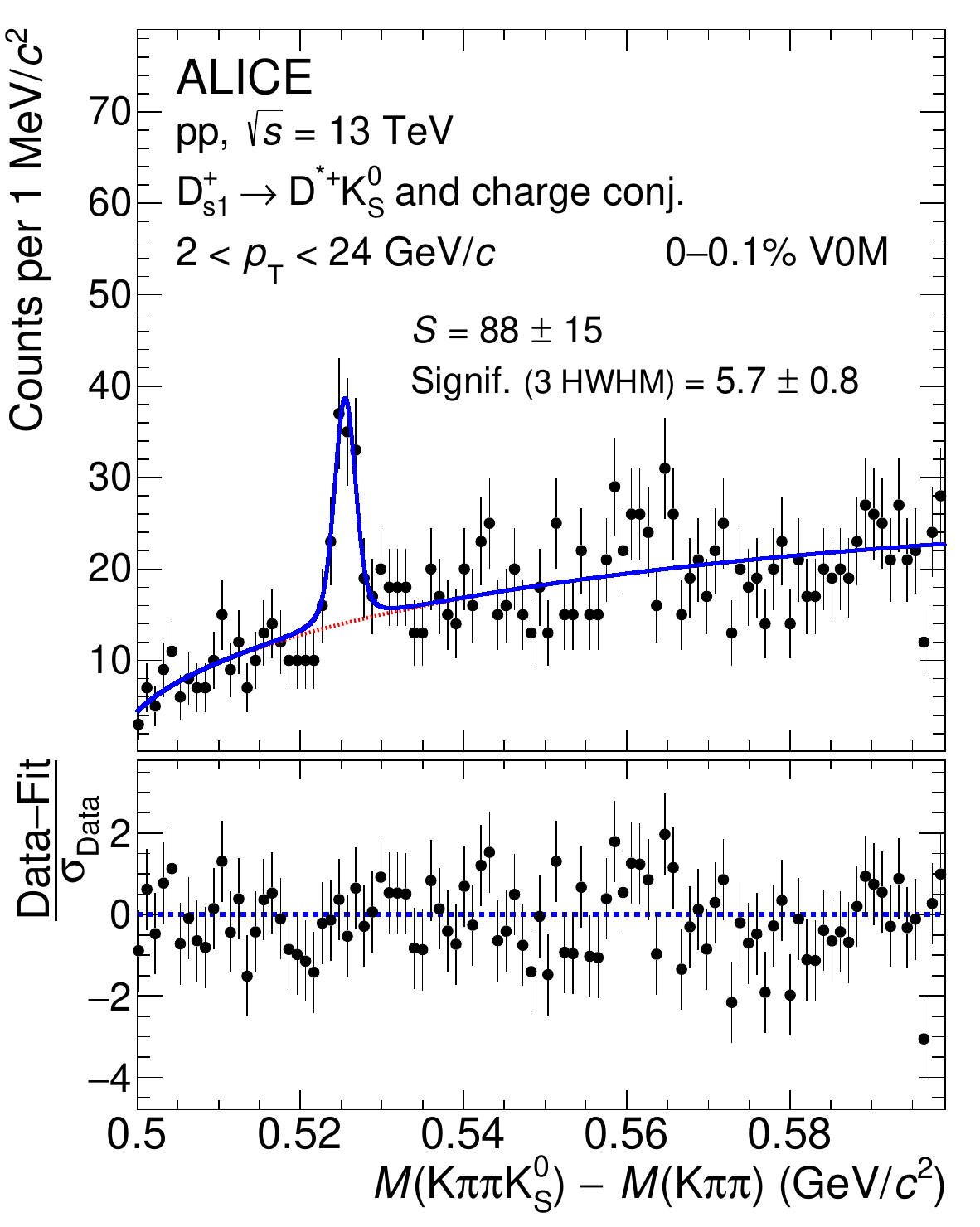}
  \includegraphics[width = 0.4\textwidth]{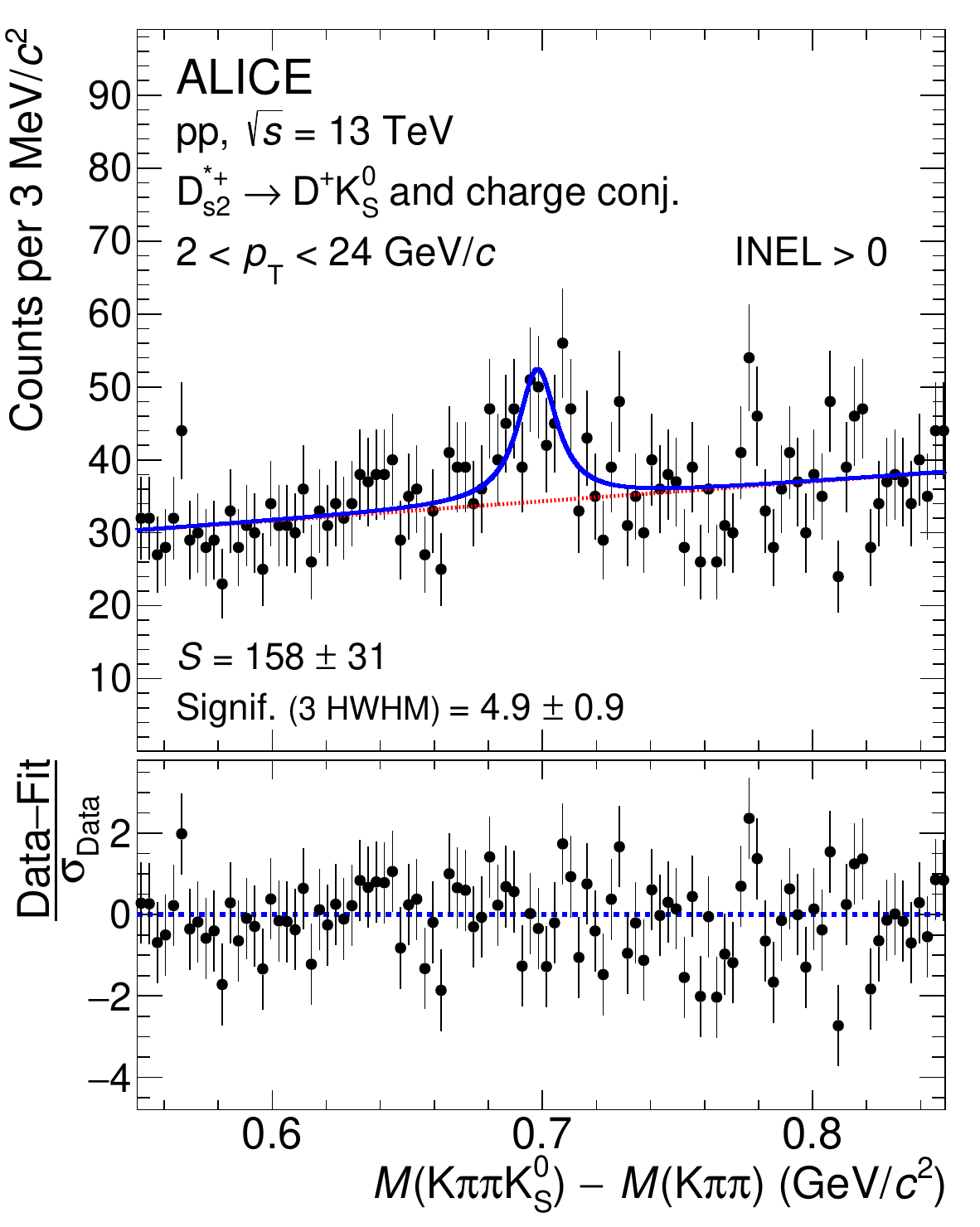}
  \includegraphics[width = 0.4\textwidth]{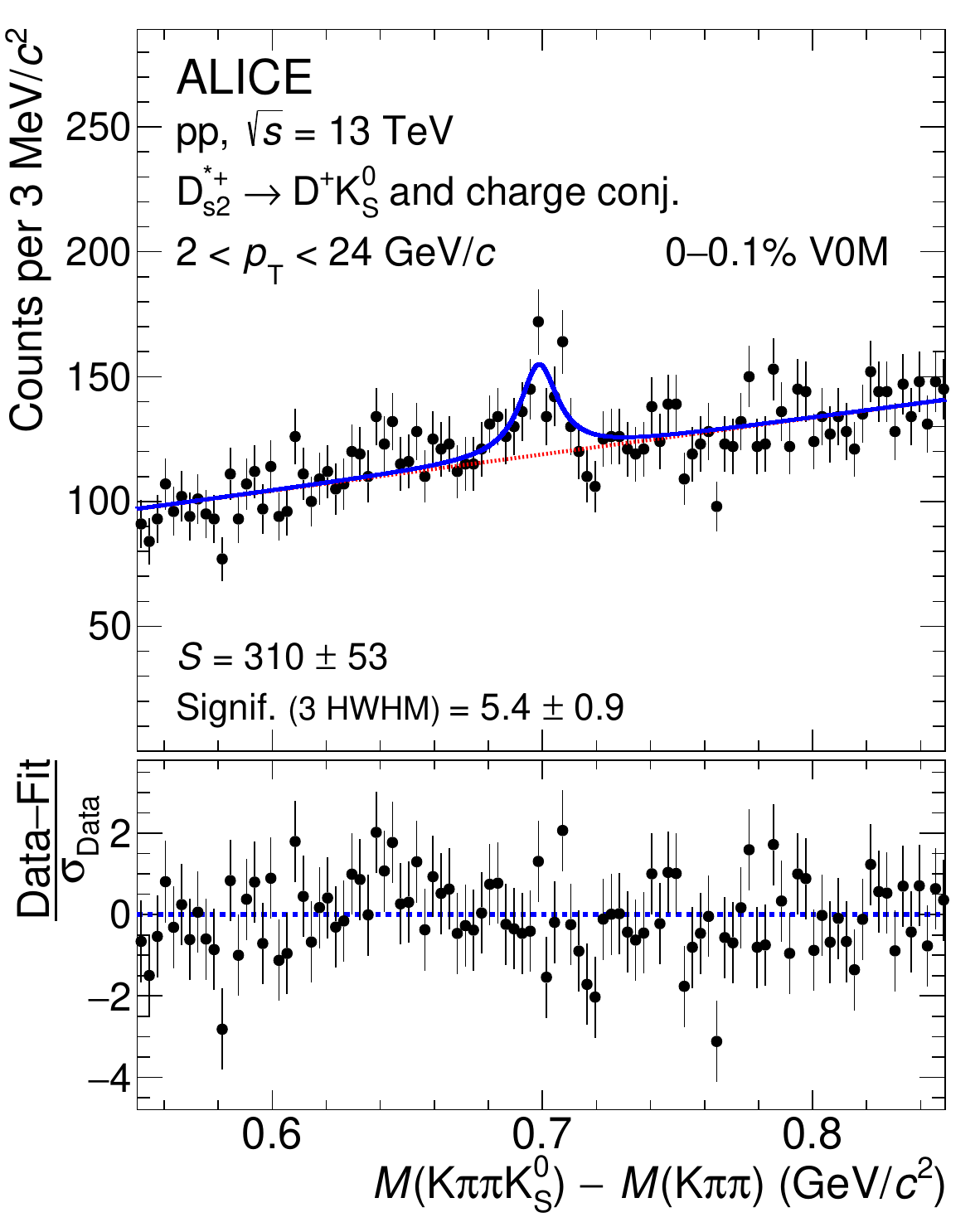}
  \caption{Invariant-mass distributions $M(\mathrm{K}\uppi\uppi \kzero) - M(\mathrm{K}\uppi\uppi)$ of $\Dsone$ (top panels) and $\Dstwo$ (bottom panels) candidates and charge conjugates in the $2<\pt<24~\GeVc$ interval, for the \INEL and HMV0 (\nolinebreak{0--0.1\% V0M}) samples. The blue solid lines show the total fit functions described in the text, and the red dashed lines represent the combinatorial background. The raw-yield ($S$) values are reported together with their statistical uncertainties, as well as the estimated significance of the signal (Signif.). The bottom sub-panels of each panel display the differences between the data and the fit, normalized by the experimental uncertainties, to show the agreement between the data and the fit.}
  \label{fig:inv_mass}
\end{figure}

The raw yields of $\Dsone$ and $\Dstwo$ mesons were computed by integrating the signal function obtained from maximum-likelihood fits to the invariant-mass distributions $\Delta M = M(\mathrm{K}\uppi\uppi \kzero) - M(\mathrm{K}\uppi\uppi)$. The raw yields were extracted in the transverse-momentum interval $2<\pt<24~\GeVc$ for both the \INEL and HMV0 samples.
The signal peak was modeled with a Voigt function, defined as the convolution of a Gaussian function and a Breit--Wigner function~\cite{grosa_2023_7579657}. The widths, $\Gamma$, of the \Ds-meson resonances were fixed to their PDG values $\Gamma(\Dsone) = (0.92 \pm 0.05)~\MeVmass$ and $\Gamma(\Dstwo) = (16.9 \pm 0.7)~\MeVmass$~\cite{ParticleDataGroup:2022pth}, while the Gaussian parameters are let free. For the $\Dsone$ meson, the background was modeled with the function \linebreak $a_0\sqrt{\Delta M - m(\kzero)}\,\mathrm{e}^{a_1[\Delta M - m(\kzero)]}$, where $m(\kzero)$ is the nominal rest mass of the \kzero meson, and $a_0$ and $a_1$ are free fit parameters. A polynomial of first order was used to describe the background in the $\Dstwo$-meson fits.
The invariant-mass distributions are reported in Fig.~\ref{fig:inv_mass} together with related fit functions. A statistically reliable signal extraction is obtained for both the \Dsone and \Dstwo mesons for \INEL and HMV0 events. The statistical significance, computed in the invariant mass region within three half-width half-maximum (HWHM) around the signal peak, ranges between 4.0 and 5.7 depending on the particle and multiplicity interval.

The corrected per-event yields times BR of prompt \Dsone and \Dstwo mesons at midrapidity were computed for each multiplicity class as
\begin{equation}
  \label{eq:CorrYield}
  \frac{1}{N^{\mathrm{ ev}}} \frac{{\mathrm{ d^2}} N}{{\mathrm{ d}}\pt{\mathrm{ d}}y  } \times \mathrm{ BR} =
  \frac{1}{2} \frac{\epsilon^{\mathrm{ INEL}}}{N^{\mathrm{\rm ev}}} \frac{1}{{\mathrm{ BR(D) \times BR(\kzero)}}} \frac{\left.f_{\mathrm{ prompt}} \times N^{\mathrm{ raw}}\right|_{|y|<y_{\mathrm{ lab}}}}{\Delta y_{\mathrm{ lab}} \times \Delta \pt \times ({\mathrm{ Acc}}\times\epsilon)_{\mathrm{ prompt}}}\;,
\end{equation}
where $N^{\rm raw}$ is the raw yield, summed for particles and antiparticles, extracted in a given multiplicity class. The raw yield is divided by the prompt acceptance-times-efficiency \effprompt and multiplied by the fraction of prompt D mesons in the selected sample $f_{\mathrm{prompt}}$ to correct for the contribution of beauty-hadron decays. It is further divided by a factor of two to obtain the charge-averaged yield, by the \pt-interval width $\Delta\pt$, and by the correction factor accounting for the rapidity coverage of reconstructed excited charm-strange mesons $\Delta y_{\rm lab}$. The term $\rm BR(D) \times BR(\kzero)$ encompasses the normalisation for the decay-channel branching ratios of the \Ds-meson resonance daughters.
The factor $N^{\rm ev}$ denotes the number of recorded events in the \INEL multiplicity class. It is corrected for the fraction of \INEL events that were selected by the trigger $\epsilon^{\rm INEL}$.

No kinematic selections were applied on the reconstructed \Dsone- and \Dstwo-meson candidates. Therefore the prompt acceptance-times-efficiency corrections for \Dstar, \Dplus, and \kzero mesons were estimated as a function of \pt, $y$, and azimuthal angle from full MC simulations, in which pp collisions are simulated using the \pythia{8.243} event generator~\cite{Sjostrand:2006za, SJOSTRAND2015159}, the generated particles are propagated through the apparatus using GEANT3~\cite{Brun:1082634} reproducing the detector layout and data-taking conditions, and the reconstruction of events is performed as in real data. They were then combined, in a second step, to obtain the \effprompt factor of \Dsone and \Dstwo mesons using a fast MC simulation based on the \pythia{8.243} decayer to describe the excited charm-strange meson decay kinematics. In this case, the \Dsone and \Dstwo mesons were sampled from the measured \pt distributions of \Ds mesons in the same multiplicity classes taken from Ref.~\citenum{ALICE:2021npz} and let decay by the \pythia{8} decayer. The efficiency was then computed by evaluating it as the product of the efficiencies of the daughter particles as a function of \pt, $\eta$, and charged-particle multiplicity estimated in number of track segments reconstructed with the first two layers of the ITS. This approach was validated by estimating the efficiency corrections of \Dsone and \Dstwo mesons using a full MC simulation with a limited number of generated events. The results of the two methods were in agreement within uncertainties.

The \Ds-resonance prompt fraction \fprompt was computed from the one of \Dstar and \Dplus mesons by accounting for the decay kinematics with an approach similar to that for the ${\mathrm{ Acc}}\times\epsilon$ factor. The D-meson \fprompt was estimated with the data-driven method introduced in Ref.~\citenum{ALICE:2021mgk}. More details on the procedure and the obtained values can be found in Ref.~\citenum{ALICE:2023brx} for \Dplus mesons and Ref.~\citenum{ALICE:2022byg} for \Dstar mesons. A correction factor of $1.9
3 \pm 0.37$  was also applied to account for the higher probability of strange quarks contained in the \Dsone and \Dstwo mesons to hadronise into a beauty rather than a charm hadron, as shown in Ref.~\citenum{ALICE:2024xln}. This factor was obtained from the measurement of prompt and non-prompt strange and non-strange D mesons in pp collisions at $\s = 13~\TeV$~\cite{ALICE:2023sgl,ALICE:2024xln}. The resulting prompt fraction of \Dsone (\Dstwo) mesons is larger than 0.75 (0.85) in both the HMV0 and \INEL samples.

The systematic uncertainties on the corrected yields of the measured $\Ds$-meson resonances in both the \INEL and HMV0 samples include the following sources: (i) extraction of the raw yield, (ii) prompt-fraction
estimation, (iii) tracking and selection efficiency evaluation, (iv) sensitivity of the efficiencies to the meson \pt shape generated in the simulation and (v) to the description of the charged-particle multiplicity. In addition, an overall normalization systematic uncertainty due to uncertainties in the BR and the integrated luminosity was considered.

The systematic uncertainty on the raw-yield extraction was assessed by repeating the yield extraction after varying the fit configuration, in particular the fit ranges, the fit function used to describe the background, and the value of the $\Gamma$ parameter by accounting for its uncertainty as reported in Ref.~\citenum{ParticleDataGroup:2022pth}. The systematic uncertainties related to the estimation of the prompt resonance fraction in the extracted yields were computed by varying the reference value of the strange-to-non-strange D-meson ratio within the range of its uncertainties. The magnitude of these uncertainties depends on the multiplicity class and ranges between 5 and 14\% for the \Dsone, and between 4 and 7\% for the \Dstwo, respectively.

The systematic uncertainty on the track-reconstruction and selection efficiencies accounts for possible discrepancies between data and MC in the ITS--TPC prolongation, track-quality selection efficiency, description of the ALICE detector material budget, and description of the topological and PID variables exploited for the \kzero and D-meson selection. The track-reconstruction systematic uncertainty was estimated by propagating the track-reconstruction systematic uncertainty on the single decay daughter track to the resonance considering the kinematic of the decay. The systematic on the selection efficiency was assessed by repeating the measurement of the \Dsone and \Dstwo corrected yields after varying the selection criteria applied to the \kzero and D mesons. The magnitude of the systematic uncertainty related to the tracking and the BDT efficiency for both \Ds states is of 6\% and 10\%, respectively.

The systematic uncertainties associated with the description of the \pt shape and the multiplicity of the events in the simulation were evaluated by varying the functional form employed to describe the \pt distribution of the generated resonances and by repeating the study employing different multiplicity weights accounting for different event-selection criteria. The assigned systematic uncertainty associated with the \pt shape varies between 10 and 13\% depending on the multiplicity class for both the \Ds states. The uncertainty associated with the multiplicity is less than 2\% for both \Ds states.

\begin{figure}[tb!]
    \centering
    \includegraphics[width=0.49\textwidth]{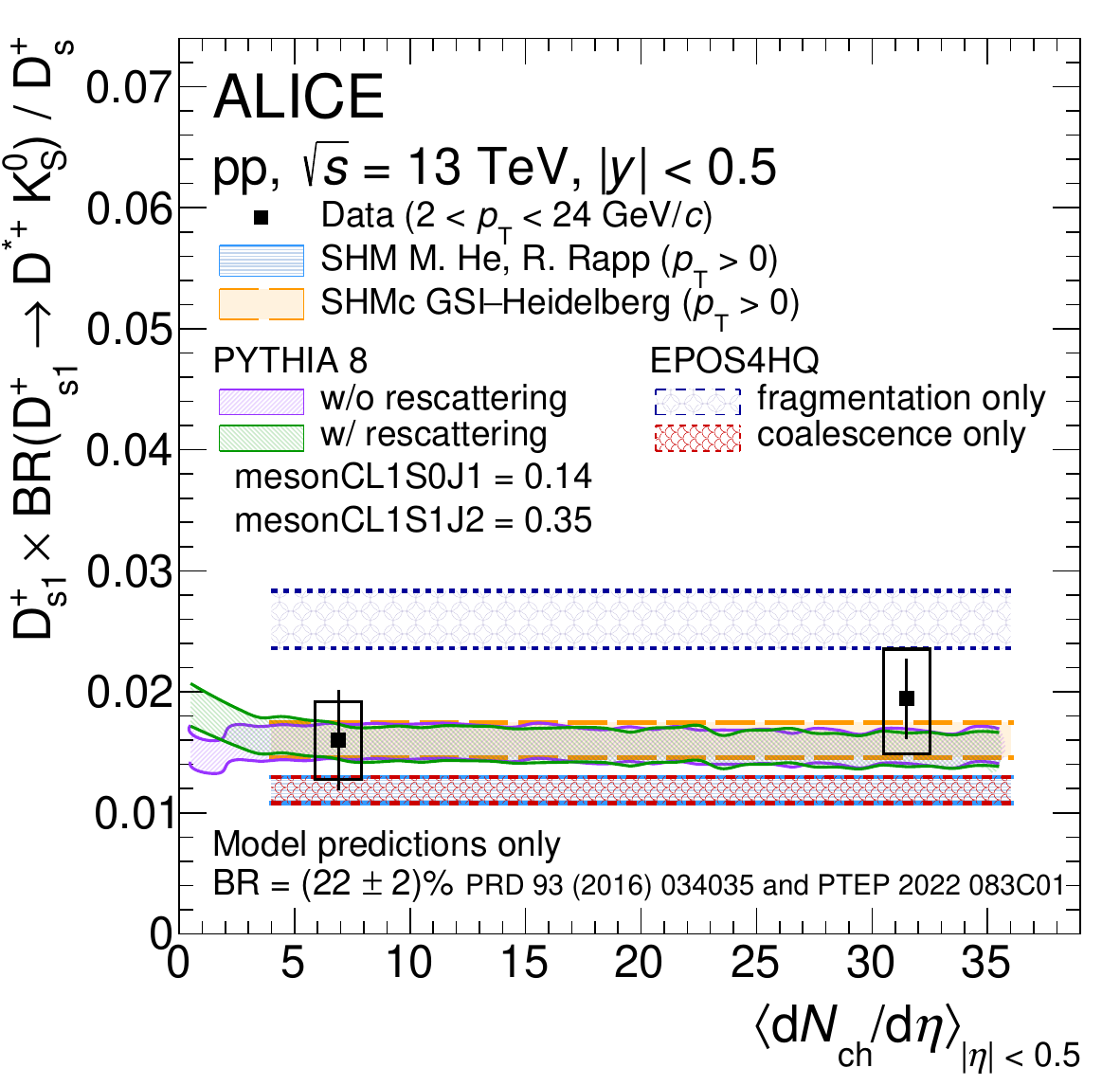}
    \includegraphics[width=0.49\textwidth]{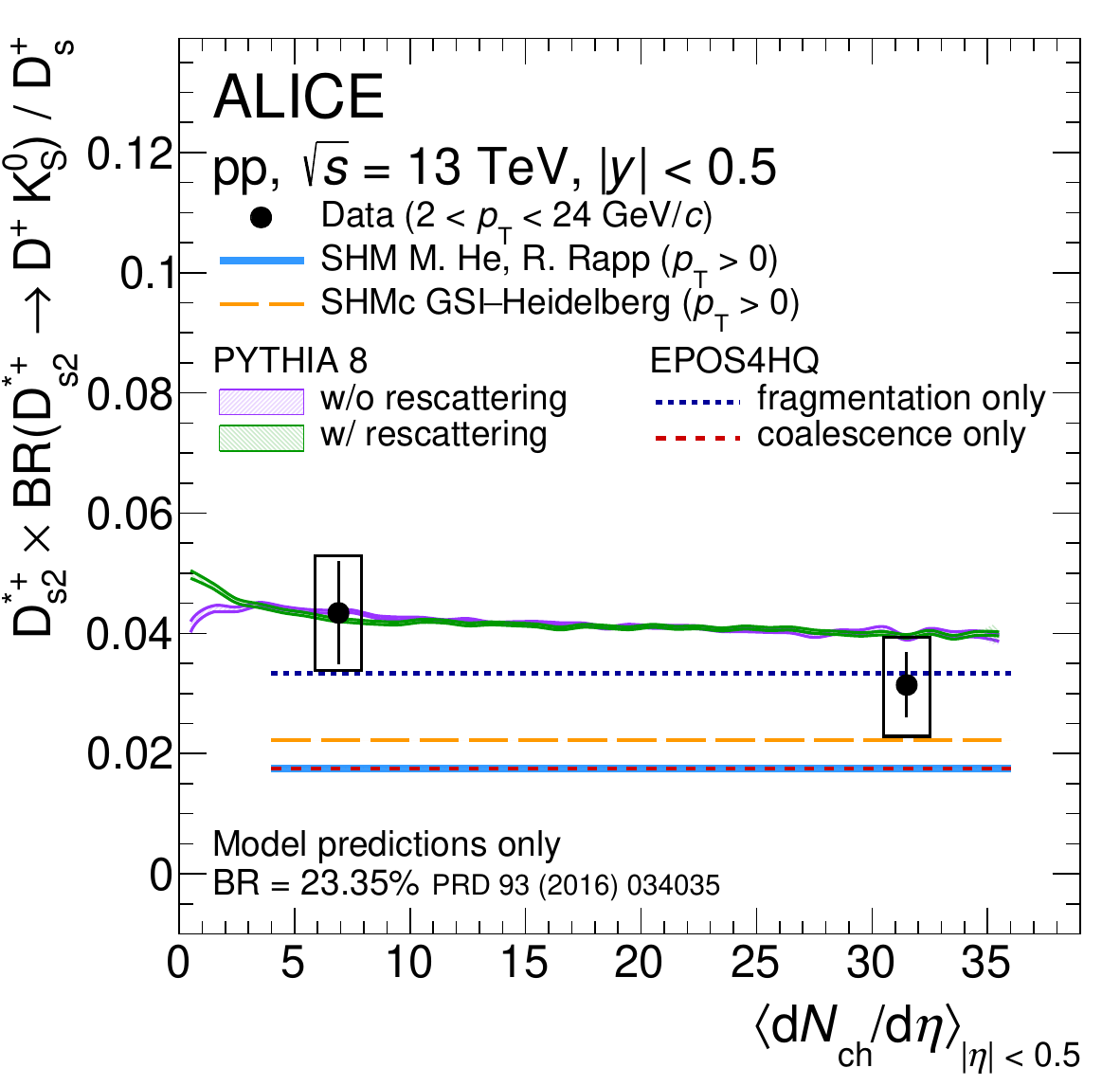}
    \caption{$\Dsone/\Ds$ (left) and $\Dstwo/\Ds$ (right) ratio times BR as a function of the average charged-particle multiplicity \avdndeta in pp collisions at $\s= 13~\TeV$ at midrapidity ($|y| < 0.5$). The experimental results are compared with the theoretical predictions based on the SHM~\cite{He:2019tik}, the SHMc~\cite{Andronic:2021erx}, PYTHIA~8~\cite{Sjostrand:2006za, SJOSTRAND2015159}, and EPOS4HQ~\cite{Zhao:2023ucp}. The statistical and systematic uncertainties on the measured ratios are depicted as vertical lines and boxes, respectively. The theoretical uncertainty on the predicted ratios is depicted as a shaded band for the \Dsone. 
    }
    \label{fig:Dsj_to_Ds_ratios}
\end{figure}

The ratios of the yields of the \Dsone and \Dstwo mesons times the relative BRs to that of the $\Ds$ meson as a function of the average charged-particle multiplicity \avdndeta at midrapidity in pp collisions at $\s = 13~\TeV$ are shown in Fig.~\ref{fig:Dsj_to_Ds_ratios}. The \pt-differential measurements of the \Ds-meson yields performed by the ALICE Collaboration for the \INEL~\cite{ ALICE:2023sgl} and HMV0~\cite{ALICE:2021npz} multiplicity classes, integrated over the \pt range between 2 and 24~GeV/$c$, were used as the denominators of the ratios. The systematic uncertainties associated with the corrected yields were treated as uncorrelated in the propagation to the ratios, except the one related to the luminosity, which was treated as fully correlated and cancels out in the ratio. The statistical and systematic uncertainties are depicted as vertical lines and empty boxes, respectively. The ratios do not show a significant dependence on \avdndeta.

The measurements were compared with the predictions based on the statistical hadronization model (SHM)~\cite{He:2019tik} and the SHM for charm hadrons (SHMc) of the GSI--Heidelberg group~\cite{Andronic:2021erx}, integrated over $\pt > 0$, and with the predictions from EPOS4HQ~\cite{Zhao:2023ucp} in two different configurations. In the SHM and SHMc, hadron abundances are dictated by thermal weights depending on the hadron rest masses. Both the SHM and SHMc include a set of not yet observed charm-baryon states. However, unlike the SHM, SHMc assumes that charm quarks are produced in initial hard scattering processes and that the total number of (anti-)charm quarks is conserved in the collision. Computing the ratio between the excited- and ground-state charm hadron yields is beneficial for the theoretical predictions as the dependency on strangeness and charm corrections cancel out exactly. The EPOS4HQ is the heavy hadron extension of the EPOS4 generator~\cite{Werner:2023zvo}. Heavy flavor quarks, produced in hard scatterings, gluon splittings, or flavor excitation, may interact with the medium constituents through elastic and inelastic processes and finally hadronise via fragmentation or coalescence. The two mechanisms are complementary and dominate at high and low momenta, respectively. The SHM, SHMc, and EPOS4HQ predictions necessitate being scaled by the BR of the analyzed hadronic decay to be fairly compared with the measurements. No measurement of these BRs is currently available, thus they were computed considering the predictions of the BR of the $\Dsone\to\mathrm{D}^{*}\mathrm{K}$ and $\Dstwo\to\mathrm{D}\mathrm{K}$ decays from the relativistic quark model (RQM)~\cite{Godfrey:2015dva} and the ratio of the BRs between the two possible final charged states. The SHM and EPOS4HQ in the pure coalescence configuration completely overlap in both the \Dsone/\Ds and \Dstwo/\Ds cases. In the case of the \Dsone, the value reported by the PDG~\cite{ParticleDataGroup:2022pth} was used with its uncertainty, while in the case of the \Dstwo an equal contribution to $\Dstwo\to\Dplus\mathrm{K^0}$ and $\Dstwo\to\Dzero\mathrm{K^+}$ was assumed, given the very similar Q-values of the decays. The resulting values are: $\mathrm{BR}(\DsonetoDstarKs) = (22 \pm 2)\%$ and $\mathrm{BR}(\DstwotoDplusKs) = 23.35\%$ for the \Dsone and \Dstwo, respectively. The uncertainties associated with the computed BR, incorporating the uncertainties on the branching fraction provided by the PDG when available, were extended to the theoretical predictions and are depicted as shaded regions for the \Dsone. The theoretical predictions from the SHM and the SHMc for the $\Dsone/\Ds$ ratio are flat as a function of multiplicity and in good agreement with the measured ratios within 0.5 and $1.2\sigma$ at high and low multiplicity, respectively. They slightly underestimate the measured central values of the $\Dstwo/\Ds$ ratio by $2\sigma$ and $1\sigma$ at low and high multiplicity, respectively. The experimental results are also compared with predictions from EPOS4HQ considering pure fragmentation or coalescence, respectively. The predicted ratios are systematically lower in the case of pure coalescence compared to pure fragmentation. A more realistic description including both the hadronization mechanisms would provide a prediction in between the ones reported in Fig.~\ref{fig:Dsj_to_Ds_ratios}, therefore more in agreement with the measured ratios. The comparison with the model predictions align with what was observed for the ratio of ground-state charm mesons in pp collisions at the LHC energies~\cite{ALICE:2021dhb, ALICE:2023sgl} indicating that a quantitative description of the relative production of charm meson states is successfully achieved using a statistical approach.

Orbitally excited states like the \Dsone and \Dstwo mesons are not considered in the PYTHIA~8 generator by default. Nevertheless, they can be included in the generation with a parameterized description of their production and subsequent decay. The parameters regulating the production of pseudovector and tensor charm mesons (i.e.~\texttt{mesonCL1S0J1} and \texttt{mesonCL1S1J2}) were tuned to minimize the $\chi^2$ between the measured excited-to-ground state ratio in the \INEL sample and the predicted one from PYTHIA~8, obtaining \texttt{mesonCL1S0J1} = 0.14 and \texttt{mesonCL1S1J2} = 0.35. In the simulation, the average charged-particle multiplicity was estimated considering the multiplicity of charged tracks at midrapidity. For the \Dsone, the uncertainty associated with the BR relative to the considered hadronic decay channel was propagated to the PYTHIA~8 predictions as done for the thermal models and is depicted as a shaded band.
In the case of the $\Dstwo/\Ds$ ratio, the data points suggest a possible decrease with increasing \avdndeta, even if the values for \INEL and HMV0 are compatible within uncertainties. To investigate the possible effect of the hadronic rescattering on the dependence of these ratios on \avdndeta, the PYTHIA~8 simulation was performed with and without enabling it~\cite{Sjostrand:2020gyg}. The two configurations produce compatible results in the considered multiplicity range. This might be related to the fact that the lifetime of the \Dstwo is longer than the expected duration of the hadronic phase in the collision and to the fact that the magnitude of hadronic interactions for D mesons with light hadrons is expected to be small in high-multiplicity pp collisions, as recently reported by the ALICE Collaboration~\cite{ALICE:2022enj, ALICE:2024bhk}. 

\begin{figure}[tb!]
    \centering
    \includegraphics[width=0.6\textwidth]{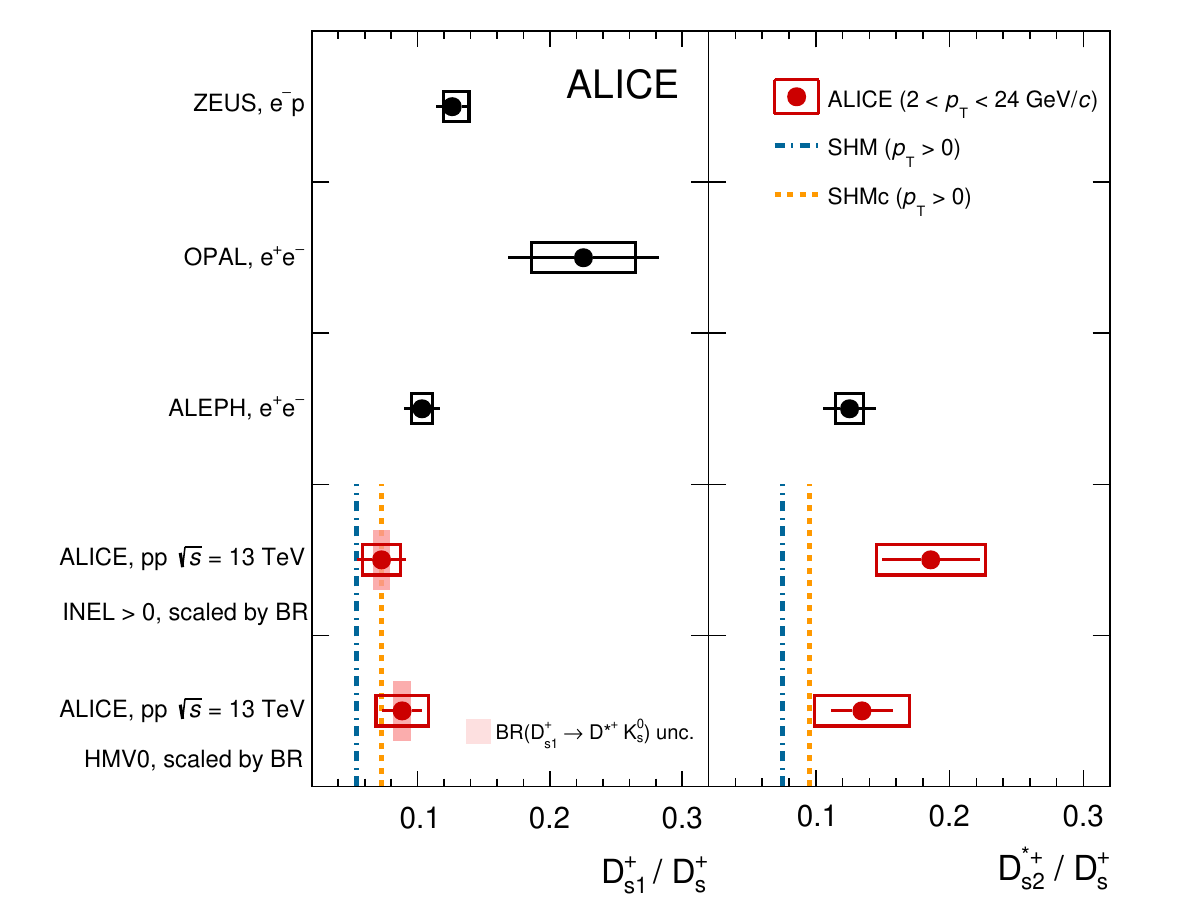}
    \caption{$\pt$-integrated yields of prompt \Dsone (left) and \Dstwo (right) mesons divided by the $\pt$-integrated yields of prompt $\Ds$ mesons. The measurements are compared to the ones performed at LEP~\cite{ZEUS:2008nzg, OPAL:1997ekz, ALEPH:2001fud, Gladilin:2014tba}. For the ALICE measurements, the results consider the BR of the decay of interest computed as described in the text considering the RQM~\cite{Godfrey:2015dva} predictions and PDG information~\cite{ParticleDataGroup:2022pth}. The experimental results are compared with the theoretical predictions based on the SHM~\cite{He:2019tik} and the SHMc~\cite{Andronic:2021erx} models.
    }
    \label{fig:Dsj_to_Ds_ratios_vs_lep}
\end{figure}

The ratios presented above can be compared with those from measurements of the fragmentation fractions of charm quarks into \Dsone, \Dstwo, and $\Ds$ mesons ($f_{\rm c}\rightarrow \rm h_{c}$) performed in \ee and \ep collisions at LEP~\cite{ZEUS:2008nzg, OPAL:1997ekz, ALEPH:2001fud, Gladilin:2014tba}. The $\Dsone/\Ds$ and $\Dstwo/\Ds$ ratios were computed as \mbox{$(f_\mathrm{c}\rightarrow \Dsone) / (f_\mathrm{c}\rightarrow \Ds)$} and $(f_\mathrm{c}\rightarrow \Dstwo) / (f_\mathrm{c}\rightarrow \Ds)$, respectively. A direct comparison of the ratios of the fragmentation fractions with the results presented in Fig.~\ref{fig:Dsj_to_Ds_ratios} cannot be made without first accounting for the BR of the decay of the resonance. The results shown in Fig.~\ref{fig:Dsj_to_Ds_ratios} were divided by the BR computed following the procedure discussed above. Figure~\ref{fig:Dsj_to_Ds_ratios_vs_lep} shows a compilation of the \Dsone/$\Ds$ and \Dstwo/$\Ds$ ratios along with the SHM predictions in pp collisions presented above. The uncertainty associated with the BR for the \Dsone is depicted separately as a shaded box. The ratios measured in pp collisions at the LHC and presented in this letter are consistent with those measured at LEP in \ee and \ep collisions with a maximum deviation of $2.1\sigma$ for \Dsone/$\Ds$ in the \INEL sample.

In summary, the production yields of \DsonePrecise and \DstwoPrecise mesons were measured for the first time at the LHC in pp collisions at $\s = 13~\TeV$ in two classes of charged-particle multiplicity. Given the absence of measurements for the BR of the considered hadronic decay channels, the results were not corrected for the relative BR. The measurements were performed at midrapidity in the $2 < \pt < 24$~GeV/$c$ range, and the ratios to the analogous measurement for the \Ds mesons scaled by the BR of the hadronic decay channel of the resonances are reported. The excited-to-ground state ratios do not show any significant dependence on the multiplicity given the current experimental uncertainty. The measurements were compared with the predictions from SHM, SHMc, and EPOS4HQ models, which quantitatively describe them.

The measured ratios were also used to tune the parameters of the \pythia{8} event generator governing the production of these orbitally excited states. In particular, these parameters were set to optimize the agreement between the measurements and the predictions from \pythia{8} at lower multiplicity. To further test the possible effects of hadronic rescattering on the excited-to-ground state ratios, the \pythia{8} simulation was performed including this effect or not. However, no significant difference was found between the two configurations.

Finally, a comparison was made between the $\Dsone/\Ds$ and $\Dstwo/\Ds$ ratios presented in this letter and the ratios of the charm-quark fragmentation fractions measured at LEP in \ee and \ep collisions. The results obtained at LHC and LEP are consistent within uncertainties. The extensive dataset collected during the Run 3 and 4 data-taking periods at the LHC will significantly reduce the relative uncertainties associated with these measurements, enabling a better understanding of the resonance production mechanism.


\newenvironment{acknowledgement}{\relax}{\relax}
\begin{acknowledgement}
\section*{Acknowledgements}

The ALICE Collaboration would like to thank all its engineers and technicians for their invaluable contributions to the construction of the experiment and the CERN accelerator teams for the outstanding performance of the LHC complex.
The ALICE Collaboration gratefully acknowledges the resources and support provided by all Grid centres and the Worldwide LHC Computing Grid (WLCG) collaboration.
The ALICE Collaboration acknowledges the following funding agencies for their support in building and running the ALICE detector:
A. I. Alikhanyan National Science Laboratory (Yerevan Physics Institute) Foundation (ANSL), State Committee of Science and World Federation of Scientists (WFS), Armenia;
Austrian Academy of Sciences, Austrian Science Fund (FWF): [M 2467-N36] and Nationalstiftung f\"{u}r Forschung, Technologie und Entwicklung, Austria;
Ministry of Communications and High Technologies, National Nuclear Research Center, Azerbaijan;
Conselho Nacional de Desenvolvimento Cient\'{\i}fico e Tecnol\'{o}gico (CNPq), Financiadora de Estudos e Projetos (Finep), Funda\c{c}\~{a}o de Amparo \`{a} Pesquisa do Estado de S\~{a}o Paulo (FAPESP) and Universidade Federal do Rio Grande do Sul (UFRGS), Brazil;
Bulgarian Ministry of Education and Science, within the National Roadmap for Research Infrastructures 2020-2027 (object CERN), Bulgaria;
Ministry of Education of China (MOEC) , Ministry of Science \& Technology of China (MSTC) and National Natural Science Foundation of China (NSFC), China;
Ministry of Science and Education and Croatian Science Foundation, Croatia;
Centro de Aplicaciones Tecnol\'{o}gicas y Desarrollo Nuclear (CEADEN), Cubaenerg\'{\i}a, Cuba;
Ministry of Education, Youth and Sports of the Czech Republic, Czech Republic;
The Danish Council for Independent Research | Natural Sciences, the VILLUM FONDEN and Danish National Research Foundation (DNRF), Denmark;
Helsinki Institute of Physics (HIP), Finland;
Commissariat \`{a} l'Energie Atomique (CEA) and Institut National de Physique Nucl\'{e}aire et de Physique des Particules (IN2P3) and Centre National de la Recherche Scientifique (CNRS), France;
Bundesministerium f\"{u}r Bildung und Forschung (BMBF) and GSI Helmholtzzentrum f\"{u}r Schwerionenforschung GmbH, Germany;
General Secretariat for Research and Technology, Ministry of Education, Research and Religions, Greece;
National Research, Development and Innovation Office, Hungary;
Department of Atomic Energy Government of India (DAE), Department of Science and Technology, Government of India (DST), University Grants Commission, Government of India (UGC) and Council of Scientific and Industrial Research (CSIR), India;
National Research and Innovation Agency - BRIN, Indonesia;
Istituto Nazionale di Fisica Nucleare (INFN), Italy;
Japanese Ministry of Education, Culture, Sports, Science and Technology (MEXT) and Japan Society for the Promotion of Science (JSPS) KAKENHI, Japan;
Consejo Nacional de Ciencia (CONACYT) y Tecnolog\'{i}a, through Fondo de Cooperaci\'{o}n Internacional en Ciencia y Tecnolog\'{i}a (FONCICYT) and Direcci\'{o}n General de Asuntos del Personal Academico (DGAPA), Mexico;
Nederlandse Organisatie voor Wetenschappelijk Onderzoek (NWO), Netherlands;
The Research Council of Norway, Norway;
Pontificia Universidad Cat\'{o}lica del Per\'{u}, Peru;
Ministry of Science and Higher Education, National Science Centre and WUT ID-UB, Poland;
Korea Institute of Science and Technology Information and National Research Foundation of Korea (NRF), Republic of Korea;
Ministry of Education and Scientific Research, Institute of Atomic Physics, Ministry of Research and Innovation and Institute of Atomic Physics and Universitatea Nationala de Stiinta si Tehnologie Politehnica Bucuresti, Romania;
Ministry of Education, Science, Research and Sport of the Slovak Republic, Slovakia;
National Research Foundation of South Africa, South Africa;
Swedish Research Council (VR) and Knut \& Alice Wallenberg Foundation (KAW), Sweden;
European Organization for Nuclear Research, Switzerland;
Suranaree University of Technology (SUT), National Science and Technology Development Agency (NSTDA) and National Science, Research and Innovation Fund (NSRF via PMU-B B05F650021), Thailand;
Turkish Energy, Nuclear and Mineral Research Agency (TENMAK), Turkey;
National Academy of  Sciences of Ukraine, Ukraine;
Science and Technology Facilities Council (STFC), United Kingdom;
National Science Foundation of the United States of America (NSF) and United States Department of Energy, Office of Nuclear Physics (DOE NP), United States of America.
In addition, individual groups or members have received support from:
Czech Science Foundation (grant no. 23-07499S), Czech Republic;
FORTE project, reg.\ no.\ CZ.02.01.01/00/22\_008/0004632, Czech Republic, co-funded by the European Union, Czech Republic;
European Research Council (grant no. 950692), European Union;
ICSC - Centro Nazionale di Ricerca in High Performance Computing, Big Data and Quantum Computing, European Union - NextGenerationEU;
Academy of Finland (Center of Excellence in Quark Matter) (grant nos. 346327, 346328), Finland. 
\end{acknowledgement}

\section*{Data availability}
This manuscript has associated data in a HEPData repository at~\cite{PaperHEPData}.

\bibliographystyle{utphys}   
\bibliography{bibliography}

\newpage
\appendix

%
%

\section{The ALICE Collaboration}
\label{app:collab}
\begin{flushleft} 
\small

S.~Acharya\,\orcidlink{0000-0002-9213-5329}\,$^{\rm 127}$, 
A.~Agarwal$^{\rm 135}$, 
G.~Aglieri Rinella\,\orcidlink{0000-0002-9611-3696}\,$^{\rm 32}$, 
L.~Aglietta\,\orcidlink{0009-0003-0763-6802}\,$^{\rm 24}$, 
M.~Agnello\,\orcidlink{0000-0002-0760-5075}\,$^{\rm 29}$, 
N.~Agrawal\,\orcidlink{0000-0003-0348-9836}\,$^{\rm 25}$, 
Z.~Ahammed\,\orcidlink{0000-0001-5241-7412}\,$^{\rm 135}$, 
S.~Ahmad\,\orcidlink{0000-0003-0497-5705}\,$^{\rm 15}$, 
S.U.~Ahn\,\orcidlink{0000-0001-8847-489X}\,$^{\rm 71}$, 
I.~Ahuja\,\orcidlink{0000-0002-4417-1392}\,$^{\rm 37}$
% ...

\section*{Affiliation Notes}

$^{\rm I}$ Deceased\\
$^{\rm II}$ Also at: Max-Planck-Institut fur Physik, Munich, Germany\\
$^{\rm III}$ Also at: Italian National Agency for New Technologies, Energy and Sustainable Economic Development (ENEA), Bologna, Italy\\
$^{\rm IV}$ Also at: Dipartimento DET del Politecnico di Torino, Turin, Italy\\
$^{\rm V}$ Also at: Yildiz Technical University, Istanbul, T\"{u}rkiye\\
$^{\rm VI}$ Also at: Department of Applied Physics, Aligarh Muslim University, Aligarh, India\\
$^{\rm VII}$ Also at: Institute of Theoretical Physics, University of Wroclaw, Poland\\
$^{\rm VIII}$ Also at: An institution covered by a cooperation agreement with CERN\\

\section*{Collaboration Institutes}

$^{1}$ A.I. Alikhanyan National Science Laboratory (Yerevan Physics Institute) Foundation, Yerevan, Armenia\\
$^{2}$ AGH University of Krakow, Cracow, Poland\\
$^{3}$ Bogolyubov Institute for Theoretical Physics, National Academy of Sciences of Ukraine, Kiev, Ukraine\\
$^{4}$ Bose Institute, Department of Physics  and Centre for Astroparticle Physics and Space Science (CAPSS), Kolkata, India\\
$^{5}$ California Polytechnic State University, San Luis Obispo, California, United States\\
$^{6}$ Central China Normal University, Wuhan, China\\
$^{7}$ Centro de Aplicaciones Tecnol\'{o}gicas y Desarrollo Nuclear (CEADEN), Havana, Cuba\\
$^{8}$ Centro de Investigaci\'{o}n y de Estudios Avanzados (CINVESTAV), Mexico City and M\'{e}rida, Mexico\\
$^{9}$ Chicago State University, Chicago, Illinois, United States\\
$^{10}$ China Institute of Atomic Energy, Beijing, China\\
$^{11}$ China University of Geosciences, Wuhan, China\\
$^{12}$ Chungbuk National University, Cheongju, Republic of Korea\\
$^{13}$ Comenius University Bratislava, Faculty of Mathematics, Physics and Informatics, Bratislava, Slovak Republic\\
$^{14}$ Creighton University, Omaha, Nebraska, United States\\
$^{15}$ Department of Physics, Aligarh Muslim University, Aligarh, India\\
$^{16}$ Department of Physics, Pusan National University, Pusan, Republic of Korea\\
$^{17}$ Department of Physics, Sejong University, Seoul, Republic of Korea\\
$^{18}$ Department of Physics, University of California, Berkeley, California, United States\\
$^{19}$ Department of Physics, University of Oslo, Oslo, Norway\\
$^{20}$ Department of Physics and Technology, University of Bergen, Bergen, Norway\\
$^{21}$ Dipartimento di Fisica, Universit\`{a} di Pavia, Pavia, Italy\\
$^{22}$ Dipartimento di Fisica dell'Universit\`{a} and Sezione INFN, Cagliari, Italy\\
$^{23}$ Dipartimento di Fisica dell'Universit\`{a} and Sezione INFN, Trieste, Italy\\
$^{24}$ Dipartimento di Fisica dell'Universit\`{a} and Sezione INFN, Turin, Italy\\
$^{25}$ Dipartimento di Fisica e Astronomia dell'Universit\`{a} and Sezione INFN, Bologna, Italy\\
$^{26}$ Dipartimento di Fisica e Astronomia dell'Universit\`{a} and Sezione INFN, Catania, Italy\\
$^{27}$ Dipartimento di Fisica e Astronomia dell'Universit\`{a} and Sezione INFN, Padova, Italy\\
$^{28}$ Dipartimento di Fisica `E.R.~Caianiello' dell'Universit\`{a} and Gruppo Collegato INFN, Salerno, Italy\\
$^{29}$ Dipartimento DISAT del Politecnico and Sezione INFN, Turin, Italy\\
$^{30}$ Dipartimento di Scienze MIFT, Universit\`{a} di Messina, Messina, Italy\\
$^{31}$ Dipartimento Interateneo di Fisica `M.~Merlin' and Sezione INFN, Bari, Italy\\
$^{32}$ European Organization for Nuclear Research (CERN), Geneva, Switzerland\\
$^{33}$ Faculty of Electrical Engineering, Mechanical Engineering and Naval Architecture, University of Split, Split, Croatia\\
$^{34}$ Faculty of Engineering and Science, Western Norway University of Applied Sciences, Bergen, Norway\\
$^{35}$ Faculty of Nuclear Sciences and Physical Engineering, Czech Technical University in Prague, Prague, Czech Republic\\
$^{36}$ Faculty of Physics, Sofia University, Sofia, Bulgaria\\
$^{37}$ Faculty of Science, P.J.~\v{S}af\'{a}rik University, Ko\v{s}ice, Slovak Republic\\
$^{38}$ Frankfurt Institute for Advanced Studies, Johann Wolfgang Goethe-Universit\"{a}t Frankfurt, Frankfurt, Germany\\
$^{39}$ Fudan University, Shanghai, China\\
$^{40}$ Gangneung-Wonju National University, Gangneung, Republic of Korea\\
$^{41}$ Gauhati University, Department of Physics, Guwahati, India\\
$^{42}$ Helmholtz-Institut f\"{u}r Strahlen- und Kernphysik, Rheinische Friedrich-Wilhelms-Universit\"{a}t Bonn, Bonn, Germany\\
$^{43}$ Helsinki Institute of Physics (HIP), Helsinki, Finland\\
$^{44}$ High Energy Physics Group,  Universidad Aut\'{o}noma de Puebla, Puebla, Mexico\\
$^{45}$ Horia Hulubei National Institute of Physics and Nuclear Engineering, Bucharest, Romania\\
$^{46}$ HUN-REN Wigner Research Centre for Physics, Budapest, Hungary\\
$^{47}$ Indian Institute of Technology Bombay (IIT), Mumbai, India\\
$^{48}$ Indian Institute of Technology Indore, Indore, India\\
$^{49}$ INFN, Laboratori Nazionali di Frascati, Frascati, Italy\\
$^{50}$ INFN, Sezione di Bari, Bari, Italy\\
$^{51}$ INFN, Sezione di Bologna, Bologna, Italy\\
$^{52}$ INFN, Sezione di Cagliari, Cagliari, Italy\\
$^{53}$ INFN, Sezione di Catania, Catania, Italy\\
$^{54}$ INFN, Sezione di Padova, Padova, Italy\\
$^{55}$ INFN, Sezione di Pavia, Pavia, Italy\\
$^{56}$ INFN, Sezione di Torino, Turin, Italy\\
$^{57}$ INFN, Sezione di Trieste, Trieste, Italy\\
$^{58}$ Inha University, Incheon, Republic of Korea\\
$^{59}$ Institute for Gravitational and Subatomic Physics (GRASP), Utrecht University/Nikhef, Utrecht, Netherlands\\
$^{60}$ Institute of Experimental Physics, Slovak Academy of Sciences, Ko\v{s}ice, Slovak Republic\\
$^{61}$ Institute of Physics, Homi Bhabha National Institute, Bhubaneswar, India\\
$^{62}$ Institute of Physics of the Czech Academy of Sciences, Prague, Czech Republic\\
$^{63}$ Institute of Space Science (ISS), Bucharest, Romania\\
$^{64}$ Institut f\"{u}r Kernphysik, Johann Wolfgang Goethe-Universit\"{a}t Frankfurt, Frankfurt, Germany\\
$^{65}$ Instituto de Ciencias Nucleares, Universidad Nacional Aut\'{o}noma de M\'{e}xico, Mexico City, Mexico\\
$^{66}$ Instituto de F\'{i}sica, Universidade Federal do Rio Grande do Sul (UFRGS), Porto Alegre, Brazil\\
$^{67}$ Instituto de F\'{\i}sica, Universidad Nacional Aut\'{o}noma de M\'{e}xico, Mexico City, Mexico\\
$^{68}$ iThemba LABS, National Research Foundation, Somerset West, South Africa\\
$^{69}$ Jeonbuk National University, Jeonju, Republic of Korea\\
$^{70}$ Johann-Wolfgang-Goethe Universit\"{a}t Frankfurt Institut f\"{u}r Informatik, Fachbereich Informatik und Mathematik, Frankfurt, Germany\\
$^{71}$ Korea Institute of Science and Technology Information, Daejeon, Republic of Korea\\
$^{72}$ KTO Karatay University, Konya, Turkey\\
$^{73}$ Laboratoire de Physique Subatomique et de Cosmologie, Universit\'{e} Grenoble-Alpes, CNRS-IN2P3, Grenoble, France\\
$^{74}$ Lawrence Berkeley National Laboratory, Berkeley, California, United States\\
$^{75}$ Lund University Department of Physics, Division of Particle Physics, Lund, Sweden\\
$^{76}$ Nagasaki Institute of Applied Science, Nagasaki, Japan\\
$^{77}$ Nara Women{'}s University (NWU), Nara, Japan\\
$^{78}$ National and Kapodistrian University of Athens, School of Science, Department of Physics , Athens, Greece\\
$^{79}$ National Centre for Nuclear Research, Warsaw, Poland\\
$^{80}$ National Institute of Science Education and Research, Homi Bhabha National Institute, Jatni, India\\
$^{81}$ National Nuclear Research Center, Baku, Azerbaijan\\
$^{82}$ National Research and Innovation Agency - BRIN, Jakarta, Indonesia\\
$^{83}$ Niels Bohr Institute, University of Copenhagen, Copenhagen, Denmark\\
$^{84}$ Nikhef, National institute for subatomic physics, Amsterdam, Netherlands\\
$^{85}$ Nuclear Physics Group, STFC Daresbury Laboratory, Daresbury, United Kingdom\\
$^{86}$ Nuclear Physics Institute of the Czech Academy of Sciences, Husinec-\v{R}e\v{z}, Czech Republic\\
$^{87}$ Oak Ridge National Laboratory, Oak Ridge, Tennessee, United States\\
$^{88}$ Ohio State University, Columbus, Ohio, United States\\
$^{89}$ Physics department, Faculty of science, University of Zagreb, Zagreb, Croatia\\
$^{90}$ Physics Department, Panjab University, Chandigarh, India\\
$^{91}$ Physics Department, University of Jammu, Jammu, India\\
$^{92}$ Physics Program and International Institute for Sustainability with Knotted Chiral Meta Matter (SKCM2), Hiroshima University, Hiroshima, Japan\\
$^{93}$ Physikalisches Institut, Eberhard-Karls-Universit\"{a}t T\"{u}bingen, T\"{u}bingen, Germany\\
$^{94}$ Physikalisches Institut, Ruprecht-Karls-Universit\"{a}t Heidelberg, Heidelberg, Germany\\
$^{95}$ Physik Department, Technische Universit\"{a}t M\"{u}nchen, Munich, Germany\\
$^{96}$ Politecnico di Bari and Sezione INFN, Bari, Italy\\
$^{97}$ Research Division and ExtreMe Matter Institute EMMI, GSI Helmholtzzentrum f\"ur Schwerionenforschung GmbH, Darmstadt, Germany\\
$^{98}$ Saga University, Saga, Japan\\
$^{99}$ Saha Institute of Nuclear Physics, Homi Bhabha National Institute, Kolkata, India\\
$^{100}$ School of Physics and Astronomy, University of Birmingham, Birmingham, United Kingdom\\
$^{101}$ Secci\'{o}n F\'{\i}sica, Departamento de Ciencias, Pontificia Universidad Cat\'{o}lica del Per\'{u}, Lima, Peru\\
$^{102}$ Stefan Meyer Institut f\"{u}r Subatomare Physik (SMI), Vienna, Austria\\
$^{103}$ SUBATECH, IMT Atlantique, Nantes Universit\'{e}, CNRS-IN2P3, Nantes, France\\
$^{104}$ Sungkyunkwan University, Suwon City, Republic of Korea\\
$^{105}$ Suranaree University of Technology, Nakhon Ratchasima, Thailand\\
$^{106}$ Technical University of Ko\v{s}ice, Ko\v{s}ice, Slovak Republic\\
$^{107}$ The Henryk Niewodniczanski Institute of Nuclear Physics, Polish Academy of Sciences, Cracow, Poland\\
$^{108}$ The University of Texas at Austin, Austin, Texas, United States\\
$^{109}$ Universidad Aut\'{o}noma de Sinaloa, Culiac\'{a}n, Mexico\\
$^{110}$ Universidade de S\~{a}o Paulo (USP), S\~{a}o Paulo, Brazil\\
$^{111}$ Universidade Estadual de Campinas (UNICAMP), Campinas, Brazil\\
$^{112}$ Universidade Federal do ABC, Santo Andre, Brazil\\
$^{113}$ Universitatea Nationala de Stiinta si Tehnologie Politehnica Bucuresti, Bucharest, Romania\\
$^{114}$ University of Cape Town, Cape Town, South Africa\\
$^{115}$ University of Derby, Derby, United Kingdom\\
$^{116}$ University of Houston, Houston, Texas, United States\\
$^{117}$ University of Jyv\"{a}skyl\"{a}, Jyv\"{a}skyl\"{a}, Finland\\
$^{118}$ University of Kansas, Lawrence, Kansas, United States\\
$^{119}$ University of Liverpool, Liverpool, United Kingdom\\
$^{120}$ University of Science and Technology of China, Hefei, China\\
$^{121}$ University of South-Eastern Norway, Kongsberg, Norway\\
$^{122}$ University of Tennessee, Knoxville, Tennessee, United States\\
$^{123}$ University of the Witwatersrand, Johannesburg, South Africa\\
$^{124}$ University of Tokyo, Tokyo, Japan\\
$^{125}$ University of Tsukuba, Tsukuba, Japan\\
$^{126}$ Universit\"{a}t M\"{u}nster, Institut f\"{u}r Kernphysik, M\"{u}nster, Germany\\
$^{127}$ Universit\'{e} Clermont Auvergne, CNRS/IN2P3, LPC, Clermont-Ferrand, France\\
$^{128}$ Universit\'{e} de Lyon, CNRS/IN2P3, Institut de Physique des 2 Infinis de Lyon, Lyon, France\\
$^{129}$ Universit\'{e} de Strasbourg, CNRS, IPHC UMR 7178, F-67000 Strasbourg, France, Strasbourg, France\\
$^{130}$ Universit\'{e} Paris-Saclay, Centre d'Etudes de Saclay (CEA), IRFU, D\'{e}partment de Physique Nucl\'{e}aire (DPhN), Saclay, France\\
$^{131}$ Universit\'{e}  Paris-Saclay, CNRS/IN2P3, IJCLab, Orsay, France\\
$^{132}$ Universit\`{a} degli Studi di Foggia, Foggia, Italy\\
$^{133}$ Universit\`{a} del Piemonte Orientale, Vercelli, Italy\\
$^{134}$ Universit\`{a} di Brescia, Brescia, Italy\\
$^{135}$ Variable Energy Cyclotron Centre, Homi Bhabha National Institute, Kolkata, India\\
$^{136}$ Warsaw University of Technology, Warsaw, Poland\\
$^{137}$ Wayne State University, Detroit, Michigan, United States\\
$^{138}$ Yale University, New Haven, Connecticut, United States\\
$^{139}$ Yonsei University, Seoul, Republic of Korea\\
$^{140}$ Affiliated with an institute covered by a cooperation agreement with CERN\\
$^{141}$ Affiliated with an international laboratory covered by a cooperation agreement with CERN.\\

\end{flushleft} 
  
\end{document}